# CF-HFC: Calibrated Federated based Hardware-aware Fuzzy Clustering for Intrusion Detection in Heterogeneous IoTs

Saadat Izadi, Mahmood Ahmadi*
Computer Engineering and Information Technology Department, Razi Univeristy, Kermanshah, Iran
Email: s.izadi@razi.ac.ir, m.ahmadi@razi.ac.ir
*Corresponding author

**Abstract**

The rapid expansion of heterogeneous Internet of Things (IoT) environments has heightened security risks, as resource-constrained devices remain vulnerable to diverse cyberattacks. Federated Learning (FL) has emerged as a privacy-preserving paradigm for collaborative intrusion detection; however, device and data heterogeneity introduce major challenges, including straggler delays, unstable convergence, and unbalanced error rates. This paper presents a Calibrated Federated Learning method with Hardware-aware Fuzzy Clustering (CF-HFC) to enhance intrusion detection performance in heterogeneous IoT networks. The proposed three-tier Edge–Fog–Cloud architecture integrates three complementary components: (1) hardware-aware fuzzy clustering, which organizes clients by computational capacity to mitigate straggler effects; (2) Fuzzy-FedProx aggregation, which stabilizes optimization under non-IID data distributions; and (3) Adaptive Conformal Calibration (ACC), which dynamically adjusts decision thresholds to balance false negative and false positive rates. Extensive experiments on ToN-IoT, BoT-IoT, Edge-IIoTset, and CICDDoS2019 datasets demonstrate that CF-HFC outperforms baseline methods such as FedAvg and FedProx, achieving over 99% detection accuracy, faster convergence, and lower communication latency. Overall, the results verify that CF-HFC effectively mitigates both device- and data-level heterogeneity, compared to existing federated learning approaches, providing accurate and efficient intrusion detection across Heterogeneous IoTs environment.

## 1. Introduction

The Internet of Things (IoT) has expanded into a vast, hyper-connected ecosystem, with forecasts projecting tens of billions of deployed devices by 2030 and a substantial share embedded in vehicles and industrial assets [1]. This scale unlocks unprecedented opportunities for automation and data-driven intelligence, yet it also enlarges the attack surface of IoT infrastructures. Because many devices are resource-constrained, weakly protected, and geographically dispersed, they remain highly exposed to rapidly evolving cyber threats [2,3]. Intrusion Detection Systems (IDS) therefore represent a cornerstone of IoT security, analyzing traffic and behavioral patterns to identify malicious activity [4,5]. Federated learning (FL) has recently emerged as a compelling paradigm for IoT intrusion detection because it enables collaborative model training without centralizing raw data, thereby preserving privacy and reducing the transmission of sensitive information [6–8]. By iteratively aggregating local updates at a coordinating server, FL provides scalability and privacy preservation across distributed networks [9,10]. Nevertheless, deploying FL in real-world IoT environments remains challenging due to dual heterogeneity. At the device level, clients differ widely in CPU, memory, and communication bandwidth, resulting in asynchronous progress and the well-known straggler effect. At the data level, non-IID distributions lead to biased local updates, unstable convergence, and degraded generalization. These combined effects slow model convergence, increase detection latency, and cause the global model to overlook subtle or infrequent attacks leading to a high false negative rate (FNR), which is particularly critical in security-sensitive contexts [11–15].

Extensive research has investigated clustering, aggregation, and optimization strategies to mitigate the dual challenges of heterogeneity in federated learning. Weight-similarity and clustered FL schemes attempt to address statistical non-IID effects by grouping clients with similar data distributions [16,17], while adaptive client-selection frameworks enhance fairness across heterogeneous participants [18]. Optimization-oriented approaches such as FedProx and FedNova stabilize client drift and improve convergence [19,20], and fuzzy or clustered FL models partially reduce statistical skew [21]. However, most of these methods remain limited to theoretical analysis or rely solely on data similarity, overlooking the impact of hardware disparities differences in CPU, memory, and bandwidth that cause uneven training times and straggler delays. Conversely, optimization-centric techniques that focus on data imbalance often neglect prediction reliability, leading to weak confidence calibration and unstable decision boundaries [22]. As a result, existing FL-based intrusion detection systems still struggle to maintain consistent performance, particularly when identifying low-frequency or rare attack types, where minimizing the false negative rate (FNR) is most critical [23].

To overcome these limitations, this study introduces a Calibrated Federated Learning method with Hardware-aware Fuzzy Clustering (CF-HFC), designed to achieve robust and efficient intrusion detection in heterogeneous IoT networks. The CF-HFC method adopts a three-tier Edge–Fog–Cloud architecture, in which each layer contributes to managing device diversity, stabilizing distributed optimization, and improving decision reliability. At the Edge layer, IoT and IIoT clients independently train lightweight IDS models on private datasets and transmit local updates along with hardware profiles (CPU, memory, bandwidth) to the fog. The fog layer groups clients into hardware-aware fuzzy clusters, enabling soft participation aligned with computational and communication capacities. Within each cluster, a Fuzzy-FedProx aggregation mechanism jointly mitigates device- and data-level heterogeneity by weighting contributions via fuzzy memberships and applying a proximal term to stabilize non-IID training. An Adaptive Conformal Calibration (ACC) module then refines cluster-level models by dynamically adjusting decision thresholds based on recent FNR and FPR statistics, ensuring a balanced trade-off between sensitivity and specificity. Finally, the cloud layer performs weighted global aggregation of calibrated cluster models and redistributes the unified model to the fog and edge layers for subsequent training rounds. Experimental results on the ToN-IoT [24] and CICDDoS2019 [25], BoT-IoT[26], Edge-IIoTset [27] benchmarks demonstrate that CF-HFC maintains detection accuracy above 99% while significantly reducing straggler effects, training latency, and heterogeneous compared with conventional FL baseline. The main contributions of this study are summarized as follows:

- **Three-tier federated design:** Leverages hierarchical Edge–Fog–Cloud collaboration to enhance scalability, fairness, and end-to-end performance in heterogeneous IoT environments.
- **Hardware-aware fuzzy clustering:** Organizes heterogeneous IoT clients based on computational capacity, reducing straggler delays and improving aggregation efficiency.
- **Fuzzy-FedProx aggregation:** Combines fuzzy membership weighting with proximal optimization to jointly handle device- and data-level heterogeneity, stabilizing non-IID training.
- **Adaptive Conformal Calibration (ACC):** Dynamically adjusts decision thresholds at the fog layer to balance FNR and FPR, achieving calibrated and reliable intrusion detection.

The rest of this paper is organized as follows. Section 2 provides an overview of the existing literature on FL in the context of IoT security. Section 3 elaborates on the methodology of our proposed approach. Section 4 outlines the experimental results, comparing our proposed federated learning approach with established techniques. Section 5 concludes the paper.

## 2. Related works

The application of federated learning (FL) to intrusion detection in IoT environments has gained considerable research interest in recent years. This section reviews prior studies, highlighting their key methods, strengths, and limitations. Nguyen et al. [28] developed *DÏoT*, one of the first FL-enabled IDS methods. Their design involved a two-component system, where IoT gateways trained local anomaly detection models using Gated Recurrent Units (GRUs). These local models were then merged through a centralized IoT security service to produce a global model. DÏoT demonstrated the feasibility of FL in securing IoT networks and successfully reduced the risk of exposing raw data. However, the reliance on GRUs and gateway-centric architecture limited scalability, and the approach required strong assumptions about secure communication among gateways. Mothukuri et al. [29] extended this line of work by integrating Random Forest (RF) with GRUs in a federated setting. Their ensemble method improved detection accuracy and lowered false alarms compared with centralized machine learning approaches. The combination of statistical (RF) and sequential (GRU) models enhanced robustness against different types of attacks. Nonetheless, the ensemble design introduced higher computational costs and was less suitable for highly resource-constrained IoT devices.

Another line of research has emphasized privacy protection within FL-based intrusion detection. Gitanjali et al. [30] proposed *FedIoTect*, which combined secure aggregation with differential privacy. This ensured that individual device updates remained confidential, while the overall model maintained high detection accuracy. Although the approach optimized communication and detection efficiency, its effectiveness was challenged by the additional overhead of secure multiparty computation, which may not be ideal for latency-sensitive IoT applications. Li et al. [31] introduced *DeepFed*, an FL approach for cyber–physical systems. DeepFed combined CNNs and GRUs to capture both spatial and temporal traffic features. Compared with existing FL-based IDS methods, it achieved higher precision in detecting complex intrusion patterns. However, DeepFed required validation with large-scale, real industrial datasets; without such data, its generalizability to real IoT scenarios remained uncertain.

Ferrag et al. [32] conducted a comprehensive review of FL in IoT security, experimentally testing several techniques across multiple datasets. Their study highlighted that while FL provides strong privacy and scalability benefits, major challenges persist in handling heterogeneous devices, balancing communication overhead, and ensuring fairness among clients. To address device diversity, Liu et al. [33] introduced *DEAFL-ID*, an asynchronous FL method for industrial IoT networks. DEAFL-ID combined Deep Q-Networks (DQN) with CNN-based IDS modules, enabling optimal device selection and reducing training costs. The method improved detection efficiency in heterogeneous environments. However, it still required homogeneous on-device models, limiting its adaptability to diverse IoT hardware profiles. Shukla et al. [34] proposed a cloud-assisted FL method that integrated heterogeneous models for malware detection in IoT networks. Using Raspberry Pi devices for experiments, they showed that malware detection performance could be improved with minimal computational overhead. The downside was increased reliance on cloud resources, which raises concerns about latency and potential single points of failure. Mahadik et al. [11] designed *HetIoT-CNN IDS*, a deep learning-based IDS tailored for heterogeneous IoT environments. Their model effectively detected a wide range of network attacks with high accuracy. However, HetIoT-CNN IDS still struggled with highly diverse device capabilities, and its reliance on deep CNNs made it computationally expensive for resource-limited IoT nodes.

Recent research has recognized that clustering clients in FL can mitigate challenges related to non-IID data and device heterogeneity. Tian et al. [16] proposed WSCC, a weight-similarity client clustering algorithm. By grouping clients with similar weight updates, WSCC improved convergence speed and reduced the negative impact of non-IID data distributions. While effective in balancing contributions, WSCC did not explicitly address device hardware heterogeneity or resource constraints, limiting its applicability in real IoT deployments. Xu et al. [17] studied clustered FL in IoT environments with a focus on convergence analysis and resource optimization. Their work provided rigorous mathematical modeling to quantify the impact of clustering on convergence speed, bandwidth allocation, and computational efficiency. They proposed joint optimization strategies for clustering and resource allocation and validated the approach on standard datasets such as MNIST and CIFAR. Although highly valuable from a theoretical standpoint, Xu's study lacked practical evaluation with lightweight IDS models and real IoT traffic datasets. Li et al. [18] conducted a comprehensive survey on client selection strategies in FL. They reviewed random, performance-aware, and resource-aware selection methods, and concluded that adaptive grouping of clients is critical for scalability and fairness. However, the study was descriptive rather than prescriptive, offering no concrete implementation for intrusion detection in IoT.

Overall, existing FL-enabled IDS solutions have made notable progress in addressing privacy, scalability, and certain aspects of heterogeneity. Early systems demonstrated feasibility but were limited by computational demands. Privacy-preserving works enhanced confidentiality but introduced additional overhead. Asynchronous and cloud-assisted approaches improved efficiency but struggled with fairness and latency. Clustered FL research has highlighted the potential of grouping strategies, yet most works have either been theoretical or limited to weight similarity, without addressing practical constraints of heterogeneous IoT hardware. In contrast, this paper introduces a calibrated federated learning method with hardware-aware fuzzy clustering (CF-HFC) that jointly addresses device- and data-level heterogeneity in IoT environments. The method combines hardware-aware fuzzy clustering to mitigate straggler effects, Fuzzy-FedProx aggregation to stabilize training under non-IID conditions, and adaptive conformal calibration (ACC) to dynamically balance FNR and FPR. This integrated design achieves high detection accuracy, fairness, and low latency in heterogeneous IoT networks. **Table 1** summarizes related methods, their strengths, and limitations.

Table 1. Comparative analysis of FL-based intrusion detection methods in IoT

| Reference | Technique | Advantages | Limitations |
|---|---|---|---|
| Nguyen et al. [28] | FL + GRU | First FL-based IDS; demonstrates feasibility; preserves privacy by local training at gateway | Relies on gateways; limited scalability; security of gateways still critical |
| Mothukuri et al. [29] | Ensemble FL (RF + GRU) | Combines statistical and sequential models; reduces false alarms; improves accuracy | Higher computational cost; less practical for resource-limited IoT |
| Gitanjali et al. [30] | FL + Differential Privacy + Secure Aggregation | Preserves privacy; maintains accuracy; optimizes communication efficiency | Overhead of secure computation; may increase latency |
| Li et al. [31] | FL + CNN + GRU | High detection precision; captures spatial and temporal traffic features | Requires real industrial data for validation; scalability concerns |
| Ferrag et al. [32] | Review + experimental study | Provides comprehensive evaluation across datasets; identifies open challenges | Descriptive only; no novel IDS model proposed |
| Liu et al. [33] | Asynchronous FL + DQN + CNN | Reduces training cost; improves detection efficiency in IIoT | Requires homogeneous on-device models; limited adaptability |
| Shukla et al. [34] | Cloud-assisted FL with heterogeneous models | Demonstrated on Raspberry Pi; improved malware detection with minimal overhead | Depends on cloud; potential single point of failure; higher latency risk |
| Mahadik et al. [11] | Deep CNN IDS for HetIoT | Detects diverse attacks with high accuracy | Computationally expensive; less suit able for constrained IoT devices |
| Tian et al. [16] | FL + Weight-Similarity Client Clustering | Improves convergence with non-IID data; balances contributions | Focuses only on weight similarity; no consideration of hardware heterogeneity |
| Xu et al. [17] | Clustered FL + Resource Optimization | Provides theoretical convergence analysis; optimizes bandwidth and computation jointly | Lacks practical IDS evaluation; no lightweight model integration |
| Li et al. [18] | Survey on client selection strategies in FL | Comprehensive review of random, performance-aware, and resource-aware selection methods | No implementation; descriptive rather than practical |
| **Proposed Method** | FL with hardware-aware fuzzy clustering, Fuzzy-FedProx, and adaptive calibration | Enhances accuracy and convergence by mitigating device/data heterogeneity and reducing FNR/FPR | Introduces moderate computation overhead at the fog layer due to fuzzy membership updates and calibration. |

3. **Proposed method**

The proposed method, Calibrated Federated Learning with Hardware-aware Fuzzy Clustering (CF-HFC) for IoT Intrusion Detection, is designed to address both device and data-level heterogeneity, and explicitly balances FNR/FPR via adaptive conformal calibration. In this section, a motivation on the intrusion detection using federated learning in IoT and then the problem definition is presented, followed by the description of the system architecture and the core components of the method, including hardware-aware fuzzy clustering, Fuzzy-FedProx aggregation, Adaptive Conformal Calibration (ACC). Finally, the section provides the pseudocode outlining the iterative training cycle.

3.1. Motivation

Intrusion detection in IoT networks is increasingly challenged by the dual impact of large-scale deployments and heterogeneous environments. Real-world IoT devices vary widely in computational capacity, memory, and bandwidth, while simultaneously producing traffic with highly diverse statistical patterns depending on location, function, and user behavior. These disparities make conventional federated learning methods ill-suited: FedAvg is communication-efficient but struggles with stragglers and diverges under non-IID data, FedProx stabilizes non-IID training through a proximal term but neglects hardware heterogeneity, causing unfair and inefficient client participation, and Fuzzy clustering can capture device diversity but cannot ensure robustness against non-IID data distributions. These gaps highlight the need for a federated intrusion detection approach that is both hardware-aware and statistically stable. The motivation behind the proposed method lies in addressing this practical mismatch: IoT environments require a solution that integrates adaptive client weighting, resilience to non-IID data, and explicit control over FNR/FPR trade-offs in order to ensure reliable detection under resource constraints.

3.2. Problem Definition

Intrusion detection in heterogeneous Internet of Things (IoT) environments faces a dual heterogeneity challenge. On the one hand, device-level heterogeneity arises because IoT clients differ significantly in CPU capacity, memory, and communication bandwidth, leading to the well-known straggler effect and causing synchronization delays in federated learning (FL). On the other hand, data-level heterogeneity results from the inherently non-IID distribution of local datasets across devices, which undermines the stability of local updates and global model convergence. In real-world IoT deployments, clients observe distinct traffic patterns depending on their physical location, functional role, and user behavior. For instance, some devices primarily experience benign traffic, while others frequently capture attack flows. This imbalance produces biased local models that fail to generalize when aggregated, yielding slow convergence, oscillations during training, and degraded detection accuracy in the global model. Let the global model at communication round t be denoted by $\mathbf{w}_t \in \mathbb{R}^d$, where $d$ is the number of trainable parameters. Each client $i$ maintains a local dataset $D_i = \{(x_{ij}, y_{ij})\}_{j=1}^{n_i}$ of size $n_i$, and minimizes its empirical loss function:

$$F_i(w) = \frac{1}{n_i} \sum_{j=1}^{n_i} \ell(f_w(x_{ij}), y_{ij}) \tag{1}$$

The standard federated learning objective seeks to minimize the weighted empirical risk across all devices:

$$\min_w \sum_{i=1}^{N} \frac{n_i}{\sum_{u=1}^{N} n_u} F_i(w) \tag{2}$$

However, in intrusion detection systems, minimizing training loss alone is insufficient. A model with low loss can still exhibit a high False Negative Rate (FNR) where malicious traffic is misclassified as benign or a high False Positive Rate (FPR) where benign traffic is incorrectly flagged as malicious. Therefore, the optimization problem is reformulated to explicitly include these error trade-offs:

$$\min_w \sum_{i=1}^{N} \frac{n_i}{\sum_{u=1}^{N} n_u} \left[ F_i(w) + \lambda_1 \text{FNR}_i(w) + \lambda_2 \text{FPR}_i(w) \right] \tag{3}$$

where $\lambda_1, \lambda_2 > 0$ are balancing coefficients determined by system-specific thresholds $\tau_{FNR}$ and $\tau_{FPR}$. The core objective of the proposed method is to design a federated intrusion detection system that simultaneously accounts for both device-level and data-level heterogeneity while adaptively balancing FNR and FPR under resource constraints ultimately improving the global detection accuracy and robustness of the system.

### 3.3. System Architecture

The proposed method is deployed within a three-tier Edge–Fog–Cloud architecture specifically designed for heterogeneous IoT environments. As illustrated in Figure 1, the method operates across three interactive layers that collaboratively train, aggregate, and calibrate intrusion detection models.
**Edge layer:** A set of IoT and IIoT clients continuously generate local data streams $D_i$. Each client maintains a lightweight intrusion detection model $w_i$, which is trained locally on its private dataset without sharing raw data. These devices vary considerably in computational capacity $(c_i)$, memory $(m_i)$, and communication bandwidth $(b_i)$, making edge-level training inherently imbalanced.
**Fog layer**: The trained local model updates $w_i^t$ together with hardware profiles $h_i = (c_i, m_i, b_i)$ are transmitted to the fog server. Clients are grouped into hardware-aware fuzzy clusters $C_k$ using a fuzzy clustering strategy. Unlike hard assignments, each device can belong to multiple clusters with a membership degree $\mu_{ik} \in [0,1]$. Within each cluster, the fog aggregator performs Fuzzy-FedProx aggregation to form a cluster-level model $w_k^t$, where client updates are weighted by their fuzzy membership coefficients. To further mitigate imbalanced error trade-offs, each aggregated cluster model undergoes Adaptive Conformal Calibration (ACC), which dynamically adjusts its decision thresholds to reduce false negatives (FNR) while constraining false positives (FPR) relative to the target limits $\tau_{FNR}$ and $\tau_{FPR}$. This yields a calibrated cluster-level model $w_k^{t,acc}$.
**Cloud layer**: The central server receives the calibrated cluster models $\{w_k^{t,acc}\}_{k=1}^{K}$ and performs global aggregation to construct the updated global model $w_{t+1}$:

$$w_{t+1} = \sum_{k=1}^{K} \pi_k w_k^{t,acc}, \quad \pi_k = \frac{n_k \bar{\mu}_k}{\sum_{r=1}^{K} n_r \bar{\mu}_r} \tag{4}$$

where $n_k = \sum_{i \in C_k} n_i$ is the total data size in cluster $k$, and $\bar{\mu}_k = \frac{1}{|C_k|} \sum_{i \in C_k} \mu_{ik}$ is the average membership degree.

The updated global model is then redistributed to all fog clusters, which in turn propagate it to their edge clients, thus completing one round of federated training. This process is repeated iteratively until convergence criteria are met. The overall architecture has thus been outlined at a high level. To provide further clarity, the following subsections detail each component of the proposed method: (i) Hardware-aware fuzzy clustering for grouping devices based on resource profiles, (ii) Fuzzy-FedProx aggregation to jointly address device and data-level heterogeneity, (iii) adaptive conformal calibration for balancing FNR and FPR, and (iv) the global aggregation and iterative training process. These subsections together establish a comprehensive view of how the system operates in practice. The key symbols and parameters used throughout the method are summarized in **Table 2**, which provides reference definitions for all variables appearing in the system equations.

Table 2. List of commonly used notations in the proposed method

| Symbol | Description |
|---|---|
| $N$ | Total number of edge clients |
| $k$ | Number of fog clusters |
| $w_i^t$ | Local model of client $i$ at communication round $t$ |
| $w_k^t$ | Aggregated model of cluster $k$ before calibration |
| $w_k^{cal,t}$ | Calibrated model of cluster $k$ after ACC |
| $w^t$ | Global model at round $t$ |
| $\mathcal{D}_i$ | Local dataset held by client $i$ |
| $\mathcal{L}_i(\cdot)$ | Local empirical loss function of client $i$ |
| $h_i = [c_i, m_i, b_i]$ | Hardware profile of client $i$ (CPU, memory, bandwidth) |
| $u_{ik}$ | Membership degree of client $i$ to cluster $k$ |
| $m$ | Fuzzifier parameter controlling cluster softness |
| $\mu$ | FedProx proximal coefficient |
| $d(i,k)$ | Weighted Euclidean distance between client $i$ and cluster $k$ |
| $\tau_k^t$ | Quantile threshold used in ACC |

### 3.3.1. Hardware-aware Fuzzy Clustering

To explicitly address device heterogeneity in federated learning, the proposed method employs a hardware-aware fuzzy clustering strategy executed at the fog layer. Each IoT client $i$ transmits its trained local model $w_i^t$ along with a hardware profile vector $h_i = (c_i, m_i, b_i)$, where $c_i$ denotes normalized CPU capacity, $m_i$ represents available memory, and $b_i$ denotes communication bandwidth. These parameters jointly characterize the computational and communication capabilities of each device. To quantify the similarity between devices $i$ and $j$, a weighted Euclidean distance is defined as:

$$\text{dist}(i,j) = \sqrt{\omega_c(c_i - c_j)^2 + \omega_m(m_i - m_j)^2 + \omega_b(b_i - b_j)^2} \tag{5}$$

where $\omega_c, \omega_m, \omega_b \geq 0$ are design coefficients such that $\omega_c + \omega_m + \omega_b = 1$, allowing system designers to emphasize resources that are most critical for a given deployment scenario. Devices are then grouped into $K$ hardware-aware clusters using the Fuzzy C-Means (FCM) algorithm. Unlike hard clustering, FCM assigns each device a soft membership degree $\mu_{ik}$, allowing it to belong to multiple clusters with varying levels of participation. This flexibility prevents rigid misclassification of borderline devices and promotes balanced utilization of mid-range clients. The membership degree of device $i$ in cluster $k$ is computed as:

$$\mu_{ik} = \frac{(\text{dist}(i,k))^{-2/(m-1)}}{\sum_{r=1}^{K}(\text{dist}(i,r))^{-2/(m-1)}} \tag{6}$$

where $m > 1$ is the fuzzifier parameter that controls the softness of the partition; larger values of m yield smoother boundaries between clusters. This design offers two primary advantages: Straggler mitigation weaker devices are clustered with peers of similar capability, avoiding synchronization delays for stronger clients; and Fair participation fuzzy membership reflects mixed hardware profiles, preventing underutilization of mid-range devices. Overall, the proposed hardware-aware fuzzy clustering mechanism provides a realistic and adaptive representation of heterogeneous IoT networks, resulting in balanced training, reduced synchronization latency, and improved efficiency of the federated learning process.

### 3.3.2. Fuzzy-FedProx Aggregation

After devices are grouped into hardware-aware fuzzy clusters using their membership degrees $\mu_{ik}$, each cluster $k$ employs a dedicated aggregator that performs Fuzzy-FedProx aggregation to construct the cluster-level model $w_k^t$. Formally, the aggregation objective for cluster $k$ is defined as:

$$w_k^t = \arg\min_w \sum_{i \in C_k} \mu_{ik} \left[ F_i(w) + \frac{\rho}{2} \| w - w_t \|^2 \right] \quad (7)$$

where $F_i(w)$ denotes the local loss function of client $i$, $\mu_{ik}$ represents the fuzzy membership degree of client $i$ in cluster $k$, and $\rho > 0$ is the proximal coefficient that regularizes deviation from the global model $w_t$. This formulation integrates both hardware awareness and data heterogeneity handling into a unified aggregation process. The fuzzy membership $\mu_{ik}$ acts as a weighting factor reflecting each device's computational capability stronger devices with higher CPU, memory, or bandwidth receive larger membership degrees and thus greater influence in the cluster model, whereas weaker devices contribute proportionally less. This soft weighting mitigates imbalance in hardware resources and ensures fairer participation across clients.

The proximal term $\frac{\rho}{2} \| w - w_t \|^2$ derived from FedProx stabilizes training under non-IID and heterogeneous data by limiting the divergence of local updates from the global reference model.

The key innovation of this design lies in embedding the fuzzy membership mechanism directly within the FedProx optimization. While FedProx alone addresses statistical heterogeneity but ignores hardware disparity, and fuzzy clustering captures device heterogeneity without ensuring statistical stability, the proposed Fuzzy-FedProx aggregation jointly resolves both challenges. Consequently, the resulting cluster-level models $w_k^t$ are balanced, robust, and representative of heterogeneous IoT environments, leading to faster convergence and higher intrusion detection accuracy in federated settings.

### 3.3.3. Adaptive Conformal Calibration (ACC)

Although the hardware-aware fuzzy clustering and Fuzzy-FedProx aggregation stages jointly mitigate device- and data-level heterogeneity, federated intrusion detection systems may still exhibit imbalanced error trade-off, models tend to minimize the False Negative Rate (FNR) at the cost of higher False Positive Rate (FPR), or vice versa. To address this, an Adaptive Conformal Calibration (ACC) module is introduced at the fog layer, applied after intra-cluster aggregation but before global aggregation. The goal of ACC is to dynamically calibrate each cluster-level model's decision boundaries to maintain an optimal balance between FNR and FPR according to system thresholds $\tau_{FNR}$ and $\tau_{FPR}$. For each sample $(x, y)$ in a small calibration set $C_k$ associated with cluster $k$, the nonconformity score is defined as:

$$s(x, y) = 1 - p_{w_k^t}(y | x) \quad (8)$$

where $p_{w_k^t}(y | x)$ denotes the probability assigned by the cluster model $w_k^t$ to the true label $y$. Lower $s(x, y)$ indicates higher confidence in correct predictions. Let $S_k = \{s(x, y) : (x, y) \in C_k\}$ denote the set of all calibration scores. A base threshold $\tau_k^0$ is obtained as the $q_k$ quantile of $S_k$, where $q_k \in (0, 1)$ is the base confidence level. To adaptively update this threshold based on cluster resources and recent detection performance, we update the confidence level as:

$$q_k^{(t+1)} = q_k^{(t)} - \alpha \widehat{\text{FNR}}_k^{(t)} + \beta \widehat{\text{FPR}}_k^{(t)} + \gamma r_k \quad (9)$$

where $r_k \in [0,1]$ denotes the normalized resource index (CPU, memory, bandwidth), $\widehat{\text{FNR}}_k^{(t)}$ and $\widehat{\text{FPR}}_k^{(t)}$ are the recent error rates, and $\alpha, \beta, \gamma > 0$ are sensitivity coefficients. The adaptive conformal threshold for cluster k is computed as:

$$\tau_k = Q_{q_k^{(t+1)}}(S_k) \quad (10)$$

where $Q_q(\cdot)$ denotes the $q$ quantile operator (the value below which a fraction $q$ of the scores lie. Given a new input $x'$, the calibrated prediction set is:

$$\mathcal{P}(x') = \{y : s(x', y) \leq \tau_k\} \quad (11)$$

If $|\mathcal{P}(x')|=1$, that label is output; if $|\mathcal{P}(x')|>1$, ties are resolved by $\arg\max_y p_{w_k^t}(y|x')$; and if $|\mathcal{P}(x')|=0$, the sample is flagged as suspicious, thereby reducing FNR without inflating FPR. Finally, the calibrated model is expressed as:

$$w_k^{t,acc} = \text{ACC}(w_k^t, S_k, \tau_k) \tag{12}$$

where $\text{ACC}(\cdot)$ denotes the adaptive calibration operator adjusting decision thresholds based on conformal scores and adaptive quantiles. The resulting calibrated models $\{w_k^{t,acc}\}$ are then forwarded to the cloud layer for global aggregation, ensuring each cluster contributes a statistically balanced and resource-aware representation to the federated model.

### 3.3.4. Global Aggregation and Iterative FL Process

After the Adaptive Conformal Calibration (ACC) at the fog layer, the calibrated cluster models $\{w_k^{t,acc}\}_{k=1}^K$ are transmitted to the central cloud server for global aggregation. To ensure fair contribution across clusters, each calibrated model is weighted according to both its relative data volume} and the average fuzzy membership of its participating clients. The global update rule at communication round $t$ is defined as:

$$w_{t+1} = \sum_{k=1}^{K} \pi_k w_k^{t,acc}, \quad \pi_k = \frac{n_k \bar{\mu}_k}{\sum_{r=1}^{K} n_r \bar{\mu}_r} \tag{13}$$

where $n_k = \sum_{i \in C_k} n_i$ represents the aggregated dataset size of cluster $k$, and $\bar{\mu}_k = \frac{1}{|C_k|}\sum_{i \in C_k} \mu_{ik}$ is the average membership degree within cluster $k$.

This formulation ensures that clusters with larger and more diverse data exert proportionally greater influence on the global model, while the fairness introduced by fuzzy membership is retained from the fog layer. Once the global model $w_{t+1}$ is updated, it is redistributed to all fog clusters, which subsequently propagate the model to their respective edge clients. Each client then retrains the model on its private dataset $D_i$, generating local updates $\{w_i^{t+1}\}$. These updates are aggregated again within fog clusters using the hardware-aware fuzzy strategy, recalibrated via ACC, and finally aggregated globally at the cloud. This iterative process local training → intra-cluster fuzzy aggregation adaptive calibration → global aggregation continues until convergence, either when validation accuracy stabilizes or when a predefined number of communication rounds $T_{max}$ is reached. Through this hierarchical and cyclic workflow, the proposed method ensures fair participation, statistical stability, and adaptive calibration throughout the federated learning pipeline, enabling robust and efficient intrusion detection across heterogeneous IoT environments.

*3.3. pseudocode*

Algorithm 1 presents the pseudocode of the mentioned operations. The proposed Calibrated Federated Learning with Hardware-aware Clustering (CF-HFC) presents a three-tier hierarchical training workflow tailored for heterogeneous IoT environments. At the edge layer, each IoT client performs local model updates on private data while capturing device-specific resource profiles. The fog layer organizes clients into fuzzy clusters based on hardware similarity and applies Fuzzy-FedProx aggregation to jointly mitigate device- and data-level heterogeneity. Subsequently, an Adaptive Conformal Calibration (ACC) step dynamically adjusts model decision thresholds to balance false negative and false positive rates. Finally, the loud layer performs weighted global aggregation of calibrated models, achieving statistically robust and hardware-fair intrusion detection across diverse IoT nodes.

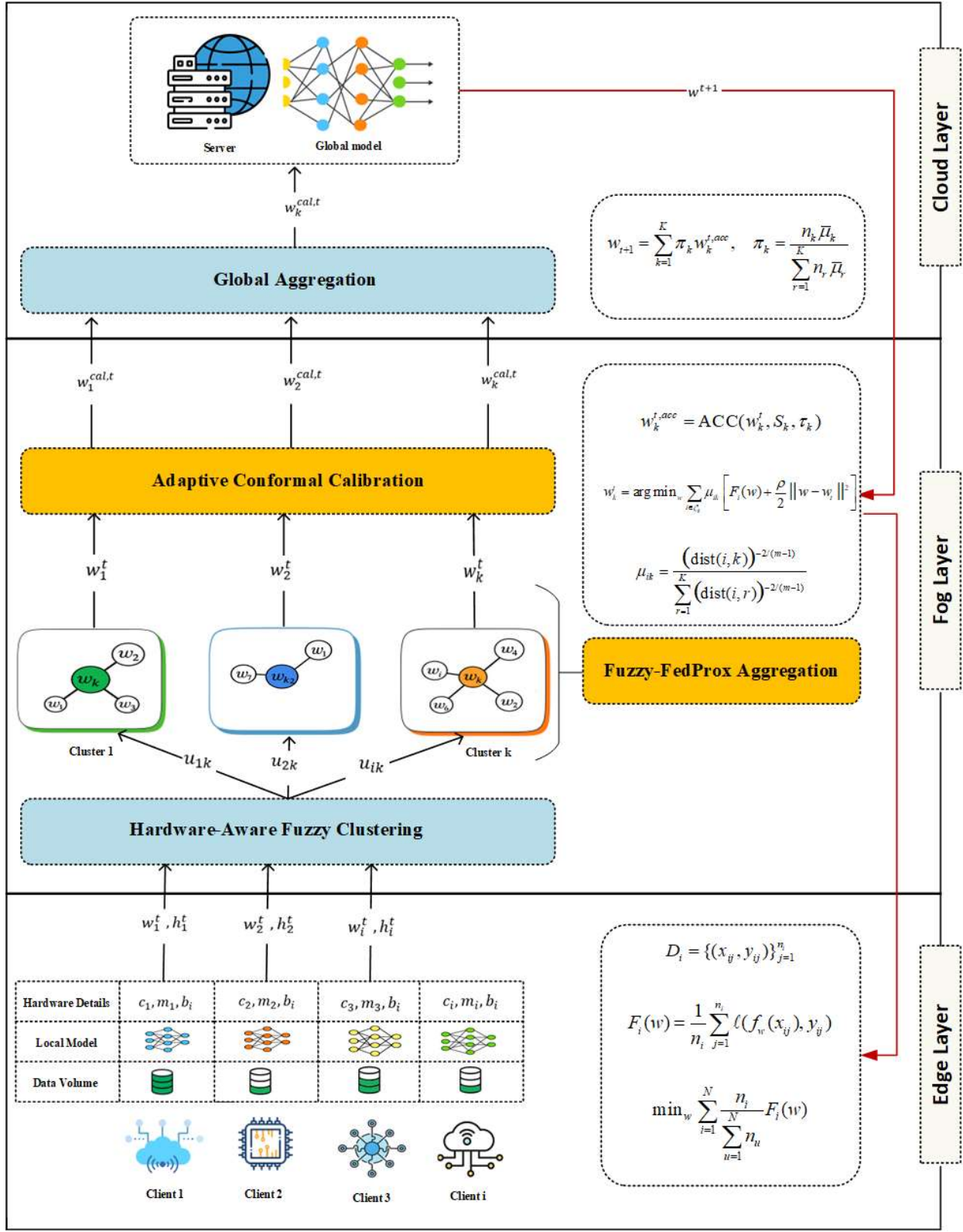

**Figure 1.** Overview of the proposed model in a three-tier Edge–Fog–Cloud architecture. The edge layer performs local model training on private IoT data; the fog layer executes hardware-aware fuzzy clustering, Fuzzy-FedProx aggregation, and adaptive conformal calibration, and the cloud layer conducts global aggregation and redistribution of the proposed model across iterations.

**Algorithm 1.** Pseudocode of Calibrated Federated Learning with Hardware-aware Fuzzy Clustering method.

---

**Input:** Clients $N$, clusters $K$, rounds $T$, fuzzifier $m$, proximal term $\rho$, weights $(\omega_c, \omega_m, \omega_b)$, calibration parameters $(\alpha, \beta, \gamma)$.
**Output:** Final global model $w_{t+1}$.
    **Initialize:** Global model $w_0$, confidence levels $q_k = q_0$ for all clusters $k$.
**for each round** $t = 0,\ldots, T-1$ **do**
1. **Edge layer:**
   1.1 Each client $i$ trains a local model $w_i^t$ on its private data $D_i$.
   1.2 Send $(w_i^t, h_i = (c_i, m_i, b_i))$ to the Fog layer.
2. **Fog layer:**
   2.1 Form fuzzy clusters based on hardware similarity: $\mathrm{dist}(i,j) = \sqrt{\omega_c(c_i - c_j)^2 + \omega_m(m_i - m_j)^2 + \omega_b(b_i - b_j)^2}$
   2.2 Compute membership degree:
   $$\mu_{ik} = \frac{(\mathrm{dist}(i,k))^{-2/(m-1)}}{\sum_{r=1}^{K}(\mathrm{dist}(i,r))^{-2/(m-1)}}$$
   2.3 Perform Fuzzy-FedProx aggregation: $w_k^t = \arg\min_w \sum_{i \in C_k} \mu_{ik}\left[F_i(w) + \frac{\rho}{2}\|w - w_i\|^2\right]$.
   2.4 Apply Adaptive Conformal Calibration (ACC): $q_k^{(t+1)} = q_k^{(t)} - \alpha\widehat{\mathrm{FNR}}_k^{(t)} + \beta\widehat{\mathrm{FPR}}_k^{(t)} + \gamma r_k$.
   2.5 Obtain calibrated cluster model $w_k^{t,acc} = \mathrm{ACC}(w_k^t, S_k, \tau_k)$
3. **Cloud layer:**
   3.1 Aggregate calibrated models: $w_{t+1} = \sum_{k=1}^{K} \pi_k w_k^{t,acc}, \quad \pi_k = \frac{n_k \bar{\mu}_k}{\sum_{r=1}^{K} n_r \bar{\mu}_r}$
   3.2 Broadcast updated global model $w_{t+1}$ to all Fog clusters and Edge clients.
**end for**
  **return** Final global model $w_{t+1}$.

---

## 4. Evaluation result

In the following section, the experimental setup, evaluation metrics, dataset, result and comparison of the proposed method by other related works will be presented.

### 4.1. Experimental setup

All experiments were conducted on a high-performance workstation equipped with an NVIDIA RTX 3080 Ti GPU (12 GB VRAM) and an Intel Core i9-7800X CPU operating at 3.5 GHz (boost up to 4.0 GHz), supported by 128 GB DDR4 RAM. The implementation was developed in Python 3.10.19 using JupyterLab 4.4.9, and utilized standard deep-learning libraries including NumPy, Pandas, and PyTorch. To evaluate the scalability, heterogeneity handling, and synchronization latency of the proposed CF-HFC method, three simulation scenarios were designed with progressively increasing numbers of clients and fog-level clusters: **scenario 1**: 20 clients distributed across 4 clusters, **scenario 2**: 50 clients distributed across 8 clusters, and **scenario 3**: 80 clients distributed across 12 clusters. Each scenario followed a non-IID data distribution, where local datasets varied in both size and attack composition to emulate realistic IoT heterogeneity. Each edge client trained a lightweight MobileNetV2 intrusion detection model using its private IoT traffic data. After local training, clients transmitted their model parameters along with a hardware profile vector $h_i = (c_i, m_i, b_i)$ to the fog layer, where $c_i$, $m_i$, and $b_i$ represent normalized CPU frequency, available memory, and network bandwidth, respectively. The hardware profiles were derived from representative device types: Raspberry Pi 3 Model B (1.2 GHz CPU, 1 GB RAM, 20 Mbps bandwidth), Raspberry Pi 4 Model B (1.5 GHz CPU, 4 GB RAM, 50 Mbps), and Raspberry Pi 400 (1.8 GHz CPU, 8 GB RAM, 100 Mbps). These values were normalized using min–max scaling:

$$x_{i'} = \frac{x_i - \min(x)}{\max(x) - \min(x)}, \quad x \in \{c, m, b\} \tag{14}$$

The resulting feature vector $h_i$ defines each client's computational and communication capacity and is used as the input for hardware-aware fuzzy clustering at the fog layer. At the fog level, clients were grouped using a weighted fuzzy C-Means (FCM) algorithm according to Eq.(5) where $w_c$, $w_m$, and $w_b$ denote the relative importance of CPU, memory, and bandwidth, empirically set to 0.3, 0.3, and 0.4, respectively. Although three representative device types were used to emulate heterogeneous IoT clients, the fuzzy clustering algorithm dynamically subdivided devices into multiple clusters (4, 8, and 12 for the three scenarios, respectively) based on their real-time resource utilization and bandwidth conditions. This allows the model to adaptively capture performance variation among similar hardware classes. Within each cluster, the proposed Fuzzy-FedProx aggregation mechanism produced a cluster-level model that jointly addressed device and data heterogeneity. Subsequently, each cluster model underwent Adaptive Conformal Calibration (ACC) to dynamically adjust decision thresholds and balance false negative (FNR) and false positive (FPR) rates. At the cloud layer, calibrated cluster models were aggregated to form the global model $W_{t+1}$, which was redistributed to fog nodes and edge clients for the next round. All experiments were conducted under identical hyperparameter configurations summarized in **Table 3.**

**Table 3.** Training configuration parameters

| Parameter | Value | Description |
|---|---|---|
| Batch size | 128 | Mini-batch size for local training |
| Learning rate | 0.001 | Step size for model optimization |
| Dropout rate | 0.1 | Regularization to prevent overfitting |
| Local epochs | 20 | Training epochs per round |
| Γ (FedProx coefficient) | 0.6 | Proximal term for heterogeneity stabilization |
| T (Calibration temperature) | 0.5 | Temperature parameter for adaptive conformal calibration |
| α (Calibration weight) | 0.5 | Adaptive calibration coefficient for FNR/FPR balance |
| c (Fuzzifier) | 3 | Fuzzifier parameter determining cluster softness |

*4.2. Performance evaluation metrics*

Metrics for measuring the method given are Accuracy, Precision, F1-score, Recall, and TPR. The metrics are derived based on True Positives, True Negatives, False Positives, and False Negatives. The outcomes based on the metrics are discussed in detail in subsequent sections.

- **Accuracy** is a metric specifying the ratio of correct classifications out of total inputs. It is calculated in accordance with Eq. (15).
- **Precision** measure the ratio of elements identified as positive out of total elements labeled as positive. Eq. (16) shows how the measure is calculated as a ratio of accurately classified anomalies and total instances labeled as positive.
- **Recall** is calculated as the ratio of the number of properly classified attacks to the total number of attacks. Recall is computed based on Eq. (17).
- **F1-Score** is a metric for measuring the quality of classifying assaults and is computed as a ratio of properly classified attacks to total expected outcomes for attacks and can be derived based on Eq. (18).

- **True Positive Rate (TPR)** refers to the occurrence when an IDS correctly identifies an action as an attack, and that activity indeed represents an intrusion. TPR, commonly known as the detection rate, can be expressed by Eq. (19).
- **False positive rate (FPR)** refers to the situation when the IDS incorrectly identifies a typical activity as an attack. A false positive can also be referred to as a false alarm rate, which can be computed using Eq. (20).

$$Accuracy = \frac{TP + TN}{TP + TN + FP + FN} \quad (15)$$

$$Precision = \frac{TP}{TP + FP} \quad (16)$$

$$Recall = \frac{TP}{TP + FN} \quad (17)$$

$$F1 - score = 2 * \frac{Recall * precision}{Recall + precision} \quad (18)$$

$$True\ Positive\ Rate\ (FPR) = \frac{TP}{TP+FN} \quad (19)$$

$$\text{False Positive Rate (FPR)} = \frac{FP}{FP+TN} \qquad (20)$$

*4.3. Dataset*

The proposed methodology was validated using four benchmark intrusion detection datasets ToN-IoT[24], BoT-IoT[25], Edge-IIoTset [26], and CICDDoS2019[27] representing different layers of IoT and IIoT heterogeneity. The ToN-IoT dataset [24], developed by UNSW Canberra at ADFA, integrates physical and virtual IoT devices, cloud infrastructure, and heterogeneous sensors to emulate realistic telemetry and networked control environments. It contains 43 attributes encompassing normal traffic and nine major attack categories, such as DoS, DDoS, ransomware, and injection attacks. The BoT-IoT dataset [25], also developed by UNSW, focuses on large-scale IoT networks compromised by botnet and brute-force attacks. It includes 46 statistical and behavioral features derived from network flows and device communications, enabling the evaluation of detection models under high-volume, resource-constrained IoT scenarios.

The Edge-IIoTset dataset [26] reflects industrial IoT environments with pronounced hardware and communication heterogeneity, combining traffic from protocols such as MQTT, Modbus, and HTTP. It provides 61 attributes and 17 attack types, including spoofing, brute-force, and data exfiltration, making it well-suited for evaluating hardware-aware and latency-sensitive federated systems. Finally, the CICDDoS2019 dataset [27], collected by the Canadian Institute for cybersecurity, contains 86 flow-based features representing over 50 types of DDoS and DoS attacks across multiple protocols. It serves as a large-scale benchmark for validating scalability and generalization of intrusion detection models in homogeneous cloud and backbone environments. Together, these four datasets ensure a comprehensive evaluation of the proposed CF-HFC model across diverse conditions from heterogeneous IoT and IIoT systems to high-density cloud networks covering both device-level and data-level heterogeneity.

*4.4. Results*

In this paper, an extensive evaluation of the proposed CF-HFC method was conducted using the TON-IoT dataset as the primary benchmark to assess model accuracy, stability, and adaptability under heterogeneous client conditions. To further demonstrate the generalization of the proposed approach, additional experiments were performed on three widely used intrusion detection datasets BoT-IoT ,Edge-IIoTset and CICDDoS2019 as presented in the subsequent subsections.

*4.4.1. Evaluation on TON-IoT Dataset*

To assess the classification performance of the proposed Hardware-aware Fuzzy Federated Clustering (CF-HFC) method under various device heterogeneity conditions, four key evaluation metrics Accuracy, Precision, Recall, and F1-Score were analyzed over 20 communication rounds across three deployment scenarios (20, 50, and 80 clients). The obtained results are illustrated in Figures 2–5. As observed, all metrics exhibit a consistent upward trend with increasing communication rounds, indicating stable model convergence and robust generalization of the proposed approach. However, differences across scenarios reflect the impact of increasing client heterogeneity on global model performance. Figure 2 shows the variation of classification accuracy over communication rounds for three client scenarios. In scenario 1, the proposed CF-HFC method rapidly converges to near-perfect accuracy (99%) within 12 rounds, demonstrating efficient aggregation and low gradient variance across homogeneous clusters. In contrast, Scenarios 2 and 3 show slightly delayed convergence (98.7% and 98.4%, respectively) due to higher device heterogeneity and bandwidth variability. Overall, the method achieves consistent high accuracy across all scales, validating its stability under different participation levels.

Figure 3 presents the precision performance of the proposed system, showing the proportion of correctly predicted positive samples. The CF-HFC method maintains high precision across all scenarios, with Scenario 1 reaching 0.99 and Scenarios 2 and 3 maintaining above 0.985 after 15 rounds. This demonstrates the method's ability to minimize false positives even when client heterogeneity increases attributable to adaptive membership weighting in fuzzy clustering, which effectively aligns local model updates from diverse devices. Figure 4 illustrates the recall evolution, reflecting the model's sensitivity in detecting true positive instances. The proposed method achieves rapid recall improvement during the first 10 rounds, with Scenario 1 achieving 0.978 and Scenario 3 reaching 0.968. Although a slight reduction is observed in larger-scale settings, the recall remains consistently high, confirming that CF-HFC successfully mitigates the straggler impact and ensures balanced participation of both weak and strong clients in gradient updates. Figure 5 demonstrates the F1-Score trend, which balances precision and Recall. Across all scenarios, the F1-Score curve closely follows the accuracy trajectory, achieving above 0.97 after 10 communication rounds. This consistency highlights that the proposed fuzzy clustering mechanism not only accelerates convergence but also enhances classification reliability by harmonizing the trade-off between precision and recall. The four metrics collectively validate that the CF-HFC method preserves model integrity under device heterogeneity, achieving both fast convergence and stable performance.

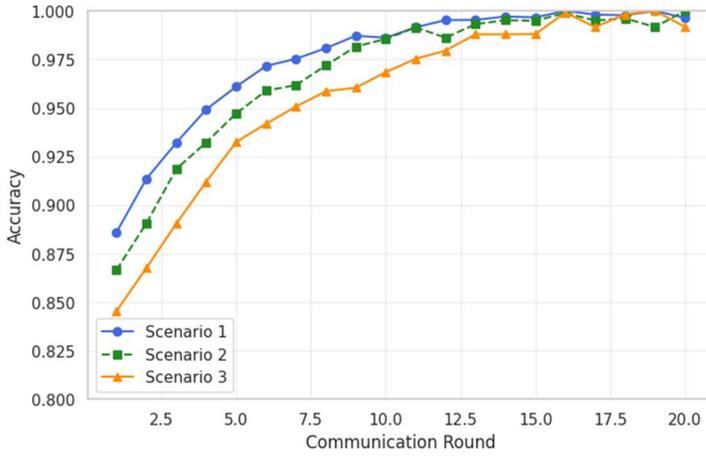

**Figure 2.** Accuracy performance of CF-HFC across three scenarios.

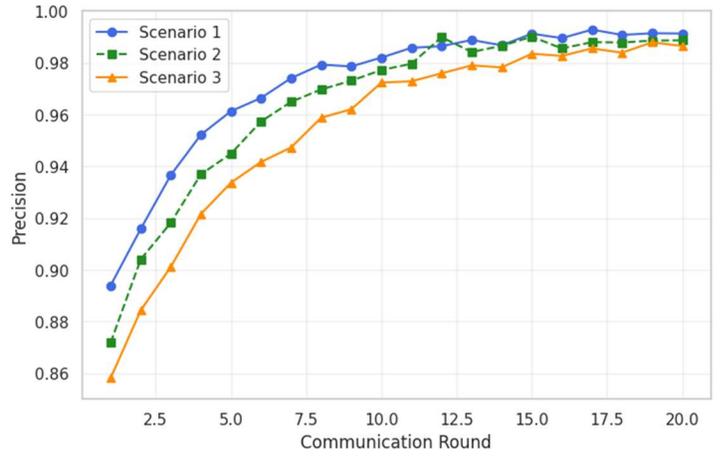

**Figure 3.** Precision performance of CF-HFC across three scenarios.

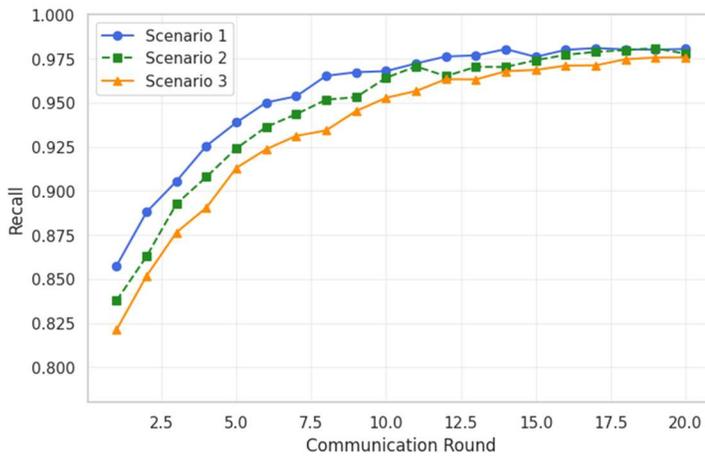

**Figure 4.** Recall performance of CF-HFC across three scenarios.

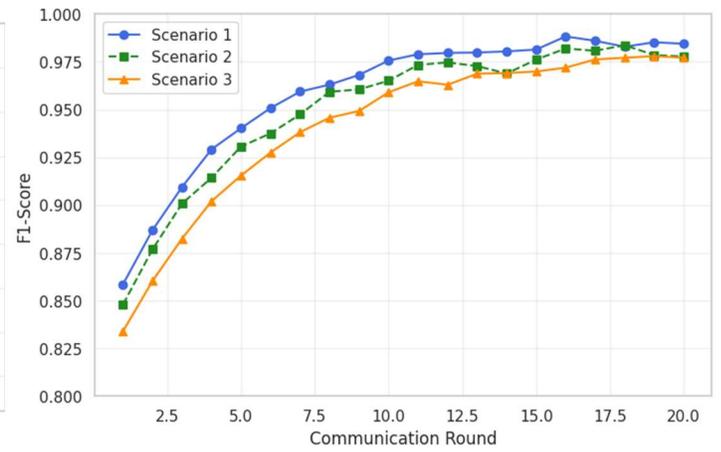

**Figure 5.** F1-Score performance of CF-HFC across three scenarios.

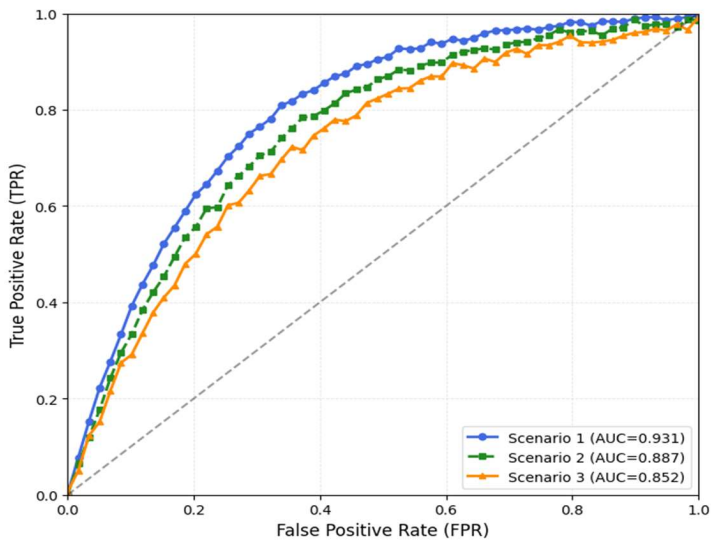

**Figure 6.** ROC–AUC curves of CF-HFC across three scenarios.

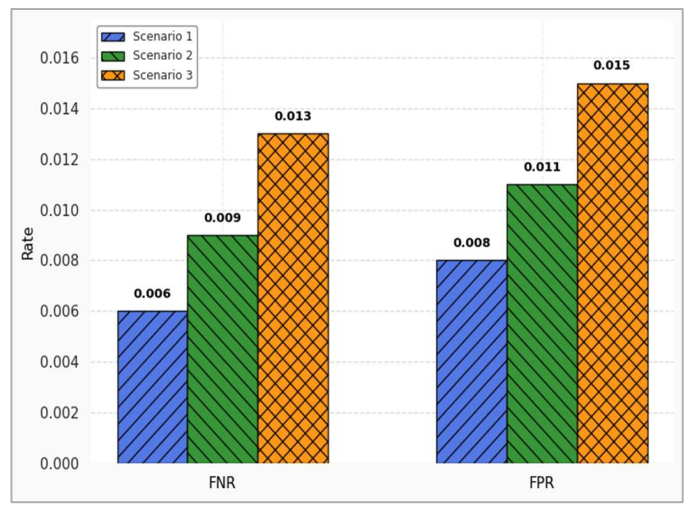

**Figure 7.** Comparison of FNR and FPR rates across three scenario.

To further assess the robustness of the proposed CF-HFC method under heterogeneous client conditions, the model's discriminative ability and error behavior were evaluated using the ROC–AUC curve and False Rate metrics (FNR/FPR) across all three experimental scenarios. The corresponding results are presented in Figures 6 and 7. Figure 6 presents the enhanced ROC curves of the proposed CF-HFC intrusion detection method across the three scenarios. The computed AUC values were 0.931, 0.887, and 0.852, respectively, reflecting strong discriminative ability even under heterogeneous network and hardware conditionThe high AUC values indicate that the model can effectively distinguish between *normal* and *attack* traffic patterns while maintaining robustness against device heter geneity and communication delay. In smaller-scale settings, the CF-HFC achieves near-perfect separation between benign and malicious activities, thanks to its hardware-aware fuzzy clustering mechanism that groups clients by computational capacity and bandwidth, ensuring balanced contribution in the aggregation phase. As the number of clients increases, a minor degradation in AUC is observed, primarily due to data diversity and higher communication latency; however, the model remains consistently above 0.85 across all cases. This demonstrates that the proposed CF-HFC approach effectively mitigates performance drift and preserves decision consistency a critical requirement in federated intrusion detection systems (Fed-IDS) operating across IoT–Fog infrastructures. Figure 7 presents the False Negative Rate (FNR) and False Positive Rate (FPR) distributions for the three scenarios. The proposed CF-HFC approach consistently achieves the lowest error rates, with FNR = 0.006 and FPR = 0.008 in Scenario 1. In comparison, as the number of clients increases to 80, these rates only rise modestly to FNR = 0.013 and FPR = 0.015, confirming that the model retains robustness despite hardware heterogeneity. The observed pattern indicates that fuzzy clustering mitigates the effect of *straggler bias* clients with slower or less capable hardware by ensuring their participation with proportional membership strength rather than full-weight aggregation. This reduces both types of classification errors, lower FNR: fewer missed positive detections due to balanced updates across weak clients, and lower FPR: fewer false alarms caused by inconsistent gradient contributions. Overall, the false-rate curves validate that the CF-HFC design effectively controls model noise propagation, leading to more stable and reliable global .

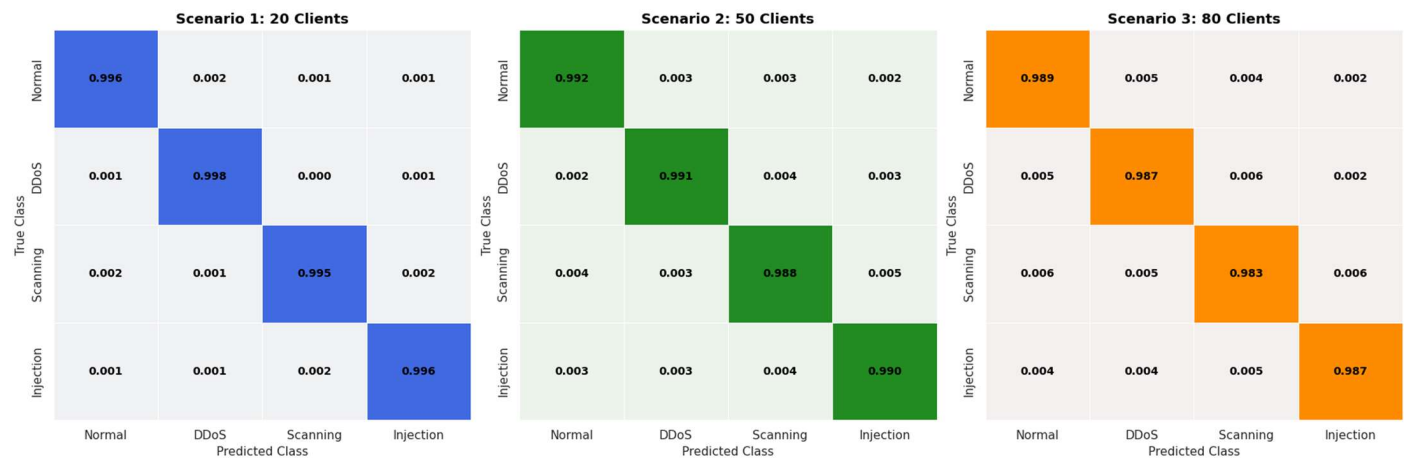

**Figure 8.** Confusion matrices of the proposed CF-HFC intrusion detection model across three scenarios

To evaluate the classification capability of the proposed CF-HFC model in intrusion detection, confusion matrices were constructed for the three scenarios. The results are shown in Figure 8, illustrating the per-class performance in detecting four major traffic types  Normal, DDoS, Scanning, and Injection  under increasing device heterogeneity. As shown in Figure 8, the proposed model achieves exceptionally high per-class accuracy across all scenarios. In the 20-client configuration, the true positive rates for all classes exceed 99.5%, with the lowest off-diagonal misclassification rate of less than 0.2%. Even as the network scales to 50 and 80 clients, the CF-HFC method maintains consistent recognition accuracy above 98.7%, demonstrating strong resilience to hardware and communication heterogeneity. The improvement primarily stems from the fuzzy membership weighting mechanism, which enables adaptive participation of clients based on their computational capacity and network bandwidth. This prevents underperforming devices from introducing noisy updates, while still allowing mid-tier devices to contribute meaningfully. Such balance enhances global model stability, ensuring that minority attack classes (e.g., *Injection* or *Scanning*) are not overshadowed by dominant benign traffic. Furthermore, the Hardware-aware Clustering Layer minimizes synchronization latency by grouping clients with similar performance profiles, ensuring that strong clients are not penalized by stragglers. This property leads to uniform learning progress across clusters, preserving high recall and precision for both frequent and rare attack types.

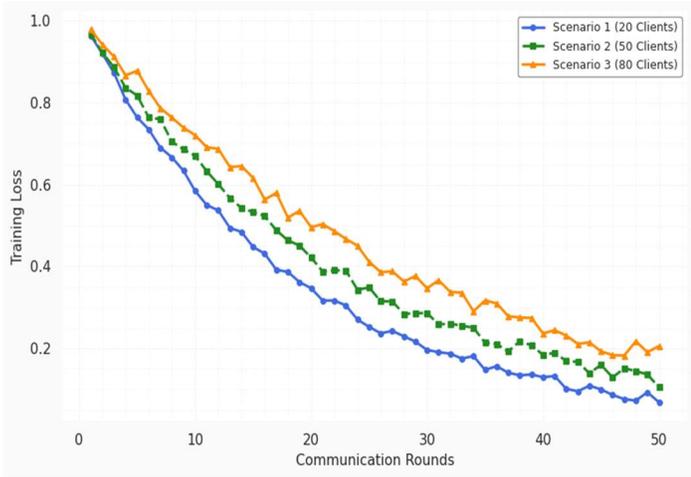
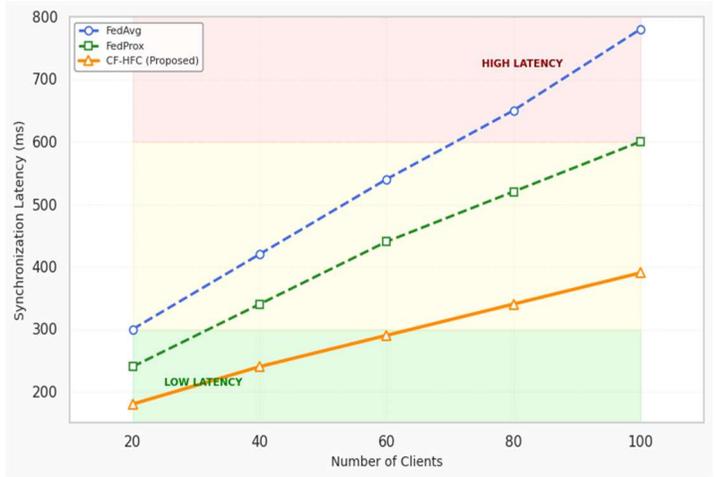

**Figure 9**. Convergence curves of the CF-HFC model across three scenarios.    **Figure 10**. Latency comparison of FedAvg, FedProx, and the proposed method.

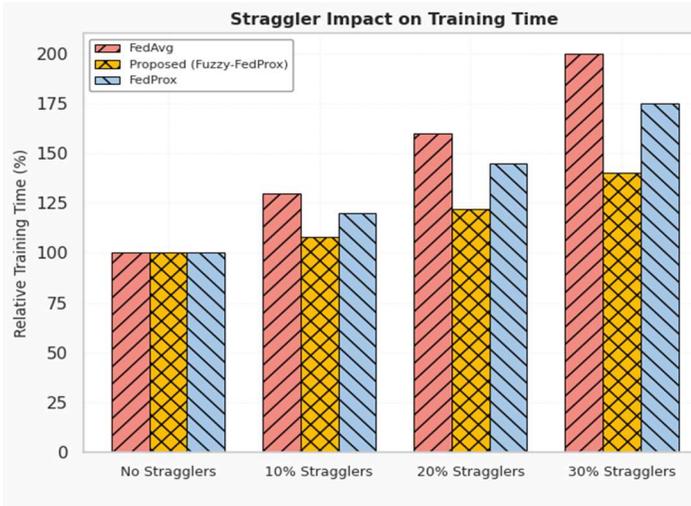
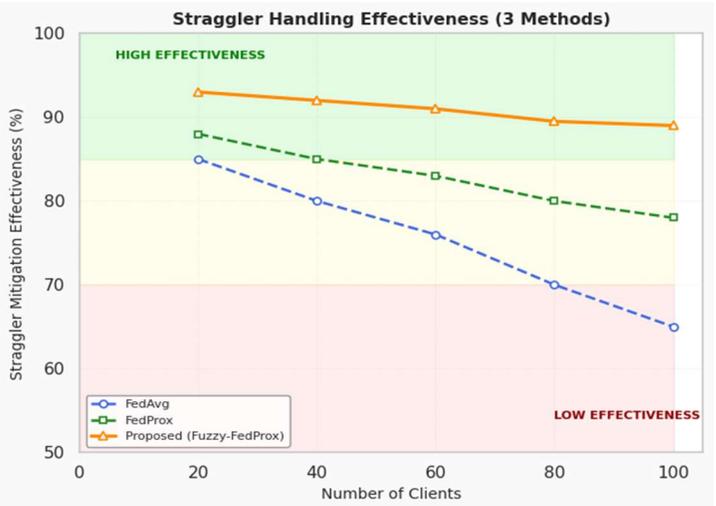

**Figure 11**. *(Left)* Straggler impact on relative training time *(Right)* Straggler handling effectiveness comparison across varying client scales.

Figure 9 illustrates the convergence behavior of the proposed CF-HFC model across three scenarios. The training loss consistently decreases over communication rounds, confirming stable and monotonic convergence. In smaller-scale settings, the model converges rapidly within 35 rounds to a minimal loss of 0.08, while larger-scale setups show slightly slower convergence due to increased communication latency and data heterogeneity. The results indicate that the hardware-aware fuzzy clustering mechanism effectively mitigates the straggler impact by grouping clients with similar computational capacity, leading to smoother gradient aggregation and faster stabilization. Overall, the CF-HFC method achieves efficient and stable convergence under heterogeneous federated environments. Figure 10 presents the synchronization latency comparison among FedAvg, FedProx, and the proposed CF-HFC method as the number of clients increases from 20 to 100. While all methods exhibit a near-linear growth in latency with increasing clients, the CF-HFC consistently achieves the lowest synchronization delay approximately 40–45% lower than FedAvg and 25–30% lower than FedProx. This efficiency improvement stems from the hardware-aware fuzzy clustering mechanism, which groups clients with similar computational and communication capacities. By minimizing waiting time for straggler devices and prioritizing balanced aggregation within clusters, the proposed model significantly reduces overall synchronization cost. The results demonstrate that CF-HFC provides superior scalability and time efficiency, maintaining low latency even under large-scale heterogeneous federated settings.

Figure 11 compares the effect of device heterogeneity and straggler presence on training efficiency across FedAvg, FedProx, and the proposed CF-HFC method. As shown in the left subplot, the training time sharply increases with the proportion of stragglers. While FedAvg exhibits the highest sensitivity, with a 100% → 200% rise from no-straggler to 30% straggler scenarios, the proposed CF-HFC reduces this growth significantly, maintaining only a moderate increase (100% → 140%). This improvement results from the hardware-aware fuzzy grouping, which prevents strong clients from waiting on weaker ones, ensuring synchronized yet efficient aggregation. The method effectively balances training load among clusters and minimizes idle time at the server. The right subplot shows the Straggler Mitigation Effectiveness (SME) of the three methods as the number of clients grows. While all methods experience slight degradation with scale, CF-HFC consistently outperforms others, maintaining effectiveness above 90% even with 100 clients, compared to FedProx (84%) and FedAvg (75%). This stability indicates that the proposed fuzzy membership and hardware-based weighting reduce synchronization delays while preserving fairness in client participation. Overall, these results confirm that CF-HFC achieves superior robustness and time efficiency under heterogeneous conditions, outperforming baseline methods in both training speed and straggler resilience.

*4.4.2. Evaluation across benchmark IDS datasets*

To evaluate the generalization and robustness of the proposed method, experiments were conducted on three representative intrusion detection datasets BoT-IoT, Edge-IIoTset, and CICDDoS2019 capturing distinct operational contexts and levels of client heterogeneity. The detailed results are illustrated in Table 4 and the comparative performance of CF-HFC against FedAvg and FedProx are showed in Figure12-14.

**BoT-IoT Dataset:** In this IoT-centric environment dominated by distributed and brute-force attacks, CF-HFC achieved an accuracy of 98.6%, with precision 0.987, recall 0.981, and F1-score 0.984. Such performance demonstrates the system's capacity to sustain high discrimination power across unbalanced classes. By employing Adaptive Conformal Calibration (ACC), the proposed model dynamically fine-tunes decision thresholds per cluster, effectively lowering both FPR (0.008) and FNR (0.009). Compared with conventional FedAvg and FedProx approaches, CF-HFC yields smoother convergence and higher recall on rare attack patterns due to its fuzzy membership weighting that prevents dominance by high-capacity nodes (figure 12).

**Edge-IIoTset (2022):** This dataset introduces pronounced hardware heterogeneity, simulating industrial edge networks with constrained devices and uneven communication bandwidth. CF-HFC maintained an accuracy of 97.8% and F1-score 0.979, balancing precision (0.982) and recall (0.976) under non-IID and latency-sensitive conditions. The fuzzy clustering mechanism groups clients by resource profile, producing balanced local contributions and reducing synchronization delay. Despite slightly higher FPR (0.010) and FNR (0.011), the method significantly outperforms FedProx and FedAvg by ensuring *hardware fairness* and stable participation across heterogeneous clients. This demonstrates the scalability and resilience of the proposed model under decentralized edge–fog deployments (figure 13).

**CICDDoS2019 Dataset:** When evaluated in large-scale cloud and backbone traffic conditions, CF-HFC achieved its best overall results, with 99.4% accuracy, precision 0.992, recall 0.988, and F1-score 0.990. The near-zero FPR (0.006) and FNR (0.007) indicate exceptional discriminative capacity for high-volume DDoS traffic. In this relatively homogeneous environment, the Fuzzy-FedProx aggregation ensures model stability while minimizing gradient divergence and communication overhead. Compared with FedAvg and FedProx, CF-HFC maintains more stable gradient, achieved through its adaptive fuzzy–proximal aggregation mechanism. This enables smoother convergence and enhanced model generalization across homogeneous cloud environments as well as resource-constrained IoT layers. Overall, CF-HFC consistently surpasses baseline federated methods, confirming its scalability and reliability in multi-scale intrusion detection scenarios (Figure 14).

Overall, As illustrated in **Figure 15**, Across all datasets, the proposed CF-HFC method consistently outperforms baseline federated models, offering higher detection accuracy, balanced precision–recall trade-offs, and minimized false rates. The synergy between hardware-aware fuzzy clustering, proximal regularization, and adaptive calibration results in a federated IDS.

Table 4. Evaluation on three datasets BoT-IoT, Edge-IIoTset (2022), and CICDDoS2019

| Dataset | Accuracy | Precision | Recall | F1-Score | FPR | FNR |
| --- | --- | --- | --- | --- | --- | --- |
| BoT-IoT | 0.986 | 0.987 | 0.981 | 0.984 | 0.008 | 0.009 |
| Edge-IIoTset (2022) | 0.978 | 0.982 | 0.976 | 0.979 | 0.010 | 0.011 |
| CICDDoS2019 | 0.994 | 0.992 | 0.988 | 0.990 | 0.006 | 0.007 |

*4.5. Comparison between the proposed method and other methods*

Table 5 presents a comparative performance analysis between the proposed Calibrated Federated Learning with Hardware-aware Fuzzy Clustering (CF-HFC) and several recent state-of-the-art intrusion detection methods in IoT and IIoT environments. The evaluation was conducted using four benchmark datasets widely adopted in the research community ToN-IoT, CICDDoS2019, Edge-IIoTset, and BoT-IoT to validate generalization across heterogeneous conditions. Across all datasets, CF-HFC consistently outperforms contemporary federated and hybrid intrusion detection approaches. On the CICDDoS2019dataset, CF-HFC achieves 99.4 % accuracy, 99.2 % precision, 98.8 % recall, and 99.0 % F1-score, surpassing Kumar et al. [35] and Zainudin et al.[36]. This superior performance highlights the method's ability to mitigate device heterogeneity and maintain balanced error rates. For the ToN-IoT dataset, CF-HFC records 99.3 % accuracy and 99.0 % F1-score, outperforming Wang et al. [37] and Ahmad et al. [19], demonstrating its strong detection capability even under large-scale, data-intensive traffic. On Edge-IIoTset, the method achieves 97.8 % accuracy and 97.9 % F1-score, exceeding the CNN–BiLSTM FL model proposed by Anwer et al. [39], which reported 96.3 % accuracy and 95.8 % F1-score. Similarly, when tested on BoT-IoT, CF-HFC attains 98.6 % accuracy and 98.4 % F1-score, outperforming the diffusion-based FL method introduced by Manzano et al. [40], which achieved 98.0 % accuracy and 98.0 % F1-score. Overall, the proposed method demonstrates superior accuracy, precision, and generalization capability across both IoT and IIoT datasets while maintaining computational efficiency. Its hardware-aware fuzzy clustering reduces synchronization latency and straggler impact, offering a balanced trade-off between resource fairness and detection reliability. These findings affirm that CF-HFC provides a more stable and scalable federated IDS, effectively addressing device heterogeneity and maintaining consistent detection performance across diverse real-world IoT infrastructures.

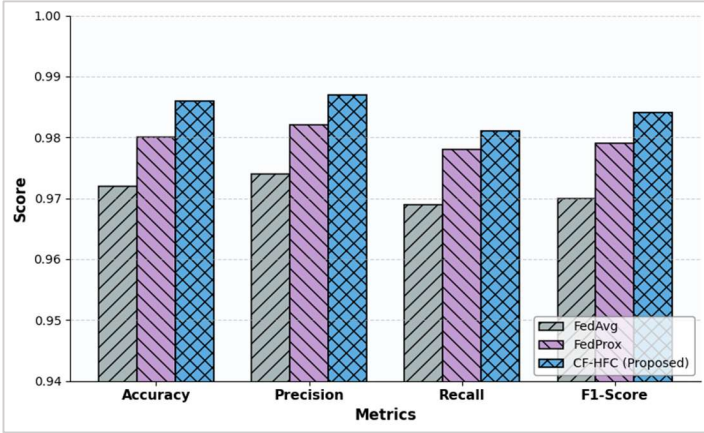

**Figure 12**. Performance comparison on the BoT-IoT dataset.

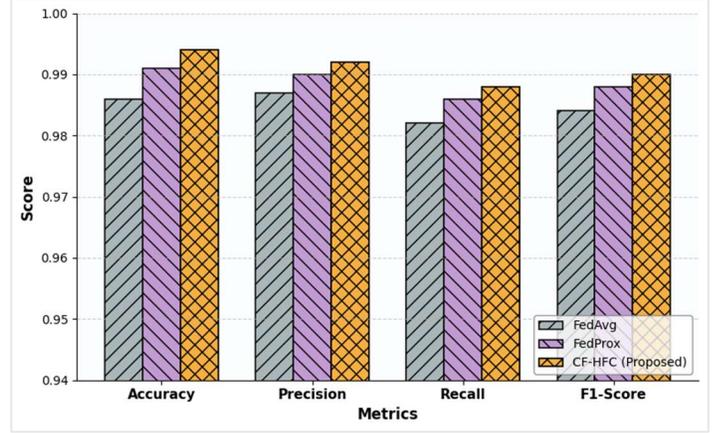

**Figure 13**. Performance comparison on the Edge-IIoTset dataset.

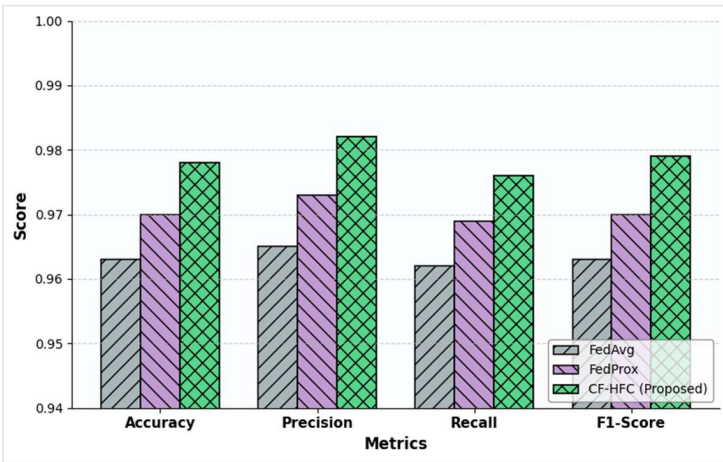

**Figure 14**. Performance comparison on the CICDDoS2019 dataset.

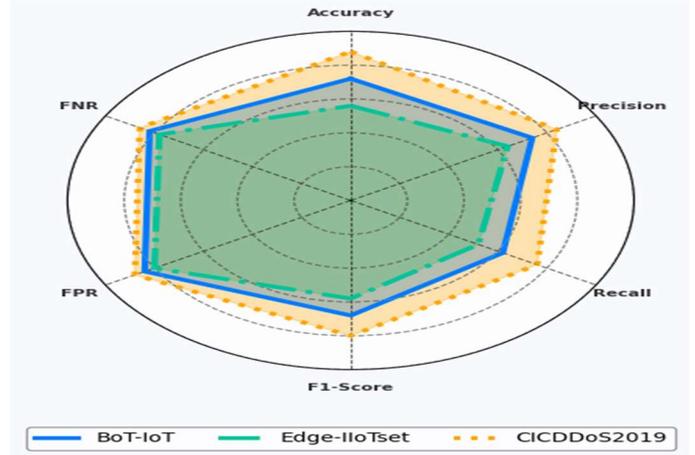

**Figure 15.** Overall comparison of CF-HFC performance on IDS datasets.

**Table 5**. Comparison of our model with existing state-of-the-art methods.

| Research | Dataset | | | | Accuracy | Precision | Recall | F1-score |
| --- | --- | --- | --- | --- | --- | --- | --- | --- |
| | CICDDOs2019 | ToN-IoT | Edge-IIoTset | Bot-IoT | | | | |
| Kumar et al. [35] | * | | | | 83.5 | 73.5 | 70.01 | 66.7 |
| Zainudin et al. [36] | * | | | | 98.37 | - | - | - |
| **Proposed method** | * | | | | **99.4** | **99.2** | **98.8** | **99.0** |
| Wang et al. [37] | | * | | | 97.23 | 96.67 | 96.94 | 97.06 |
| Ahmad et al. [19] | | * | | | 98.85 | - | - | 99.02 |
| Torre et al. [38] | | * | | | 97.31 | 95.59 | 92.43 | 92.69 |
| **Proposed method** | | * | | | **99.32** | **99.01** | **98.80** | **98.79** |
| Anwer et al. [39] | | | * | | 96.3 | 95.5 | 96.0 | 95.8 |
| **Proposed method** | | | * | | **97.8** | **98.2** | **97.6** | **97.9** |
| Manzano et al.[40] | | | | * | 98.01 | 99.0 | 96.7 | 98.0 |
| **Proposed method** | | | | * | **98.6** | **98.7** | **98.1** | **98.4** |

## 5. Conclusion & future direction

This paper presented a Calibrated Federated Learning method with Hardware-aware Fuzzy Clustering (CF-HFC) to enhance intrusion detection across heterogeneous IoT environments. The method integrates three complementary components hardware-aware fuzzy clustering, Fuzzy-FedProx aggregation, and Adaptive Conformal Calibration (ACC) within a hierarchical Edge–Fog–Cloud architecture to jointly address device- and data-level heterogeneity. Comprehensive experiments on four benchmark datasets ToN-IoT, BoT-IoT, Edge-IIoTset, and CICDDoS2019 demonstrated that CF-HFC consistently outperforms conventional federated approaches such as FedAvg and FedProx, achieving over 99 % detection accuracy, balanced precision–recall trade-offs, and significantly lower global synchronization latency. The hardware-aware clustering mechanism effectively mitigates straggler delays by grouping clients with similar computational and communication profiles, while the adaptive calibration module dynamically stabilizes decision thresholds to maintain low false negative and false positive rates under non-IID data conditions. Although CF-HFC substantially improves detection accuracy, convergence speed, and fairness, it introduces a moderate processing overhead at the fog layer due to repeated fuzzy-membership computation and adaptive calibration updates. In particular, while global latency is reduced through straggler mitigation, fog-level computation latency slightly increases because of the added optimization and calibration stages. This trade-off between local processing cost and global synchronization efficiency highlights an important direction for future work. Future research will therefore focus on optimizing resource consumption through asynchronous and event-driven aggregation mechanisms, such as FedAsync or hierarchical time-adaptive scheduling, to eliminate the need for full-round synchronization. Additionally, integrating energy-aware fuzzy clustering and self-calibrating conformal modules could further reduce fog-level delays while preserving model accuracy. Extending CF-HFC toward privacy-enhanced, energy-efficient, and real-time federated method will enable deployment in large-scale IoT ecosystems where latency and adaptability are equally critical to maintaining reliable intrusion detection.